# Investigation of electronic energy levels in a weak ferromagnetic oxygen-deficient BiFeO$_{2.85}$ thick film using absorption and X-ray photoelectron spectroscopic studies

Ramachandran Balakrishnan,[*,a,b] Ambesh Dixit[c] and Mamidanna Sri Ramachandra Rao[b]

[a]Physics Research Centre, Department of Physics, *Sethu Institute Technology, Kariapatti-626111, Tamil Nadu, India.*

[b]*Nano Functional Materials Technology Centre,* Department of Physics, *Indian Institute of Technology Madras, Chennai 600036, Tamil Nadu, India.*

[c]Advanced Materials and Devices Laboratory, *Department of Physics, Indian Institute of Technology Jodhpur, Jodhpur 342011, Rajasthan, India*

[*]Author for correspondence: ramskovil@gmail.com (BR)

**Abstract:**

We grew a 2 μm-thick film of single-phase BiFeO$_3$ on a Si (100) substrate by pulsed laser deposition with a substrate temperature of 575 °C and an oxygen partial pressure of 0.06 mbar. X-ray diffraction analysis indicated that the film exhibits textured growth along the (110) plane and possesses a rhombohedral *R3c* structure. Investigations using scanning electron microscopy and atomic force microscopy revealed an average grain size of about 300 nm and a surface roughness of 18 nm for the film. Energy-dispersive X-ray analysis estimated the composition of the film to be BiFeO$_{2.85}$. Temperature- and magnetic field-dependent magnetization measurements demonstrated weak ferromagnetic properties in the BiFeO$_{2.85}$ film, with a non-zero spontaneous magnetization at $H = 0$ Oe across the temperature range of 2–300 K. Furthermore, the exchange bias field ($H_{EB}$) of the film changed from the positive exchange bias field ($+H_{EB} = +6.45$ Oe) at 200 K to a negative field ($-H_{EB} = -8.12$ Oe) at 100 K, indicating a shift in macroscopic magnetism from antiferromagnetic to weak ferromagnetic order below 200 K. Elemental analysis via X-ray photoelectron spectroscopy revealed that the Fe ions in the BiFeO$_{2.85}$ film are in a 3+ valence state, and a peak feature at 532.1 eV confirmed the presence of induced oxygen vacancies. UV-visible-NIR and valence band spectroscopic studies showed that the direct band-gap energy, and the separation between the valence band maximum and Fermi energy were approximately 2.27 eV and 0.9 eV, respectively, which are red-shifted when compared to its bulk form.

## 1. INTRODUCTION

In the past two decades, new-generation materials such as multiferroics have attracted enormous interest among researchers in the fields of science and technology.[1] Due to the coexistence of two or more ferroic orders in multiferroic materials, they have been extensively investigated for various magnetoelectric (ME) and spintronic applications.[1-5] These applications include high-density data storage devices, high-frequency filters, magnetic sensors, transducers, actuators, and photovoltaics. Perovskite bismuth ferrite ($BiFeO_3$) is one of the many lead-free multiferroic materials studied for ME applications.[2-11] However, scientists and engineers have encountered two major issues: processing temperature and atmosphere during the growth of $BiFeO_3$ films on various substrates using different growth techniques.[12-18]

In a study by Bea et al.,[18] they discovered the occurrence of parasitic phases when growing 70 nm thick $BiFeO_3$ thin films on an STO (001) substrate using pulsed laser deposition. At a low substrate temperature ($T_S$ = 520 ºC) or high oxygen partial pressure ($O_{2pp}$ = 0.1 mbar), the impurity phase $Bi_2O_3$ formed alongside the desired phase $BiFeO_3$, while a secondary phase $Fe_2O_3$ formed at high-$T$ ($T_S$ = 750 ºC) or low $O_{2pp}$ ($10^{-4}$ mbar). Notably, Bea et al.[18] were able to successfully grow epitaxial pseudo-cubic $BiFeO_3$ thin films under intermediate growth conditions of $T_S$ = 520–750 ºC and $O_{2pp}$ = $10^{-4}$–$10^{-1}$ mbar. However, the rhombohedral structure (R-phase) is only thermodynamically stable in thick $BiFeO_3$ films or polycrystalline samples.[17] The stable R-phase $BiFeO_3$ films have great potential for practical applications in ferroelectric, piezoelectric, and optoelectronic devices. However, there is a discrepancy in the band-to-band transition of $BiFeO_3$, whether it is direct and/or indirect, and its reported band-gap values in a wide range of 0.9-2.8 eV in the literature.

Thus, in this study, we grew a BiFeO$_3$ thick film on an *n*-type Si (100) substrate under optimized growth conditions at $T_S$ = 575 ºC and $O_{2pp}$ = 0.06 mbar using a pulsed laser deposition technique. The film was then analyzed using X-ray diffraction, UV-visible-NIR spectroscopy, X-ray photoelectron spectroscopy, and magnetization measurements. Our findings indicate that the grown oxygen-deficient BiFeO$_{2.85}$ thick film is single-phase and polycrystalline with a rhombohedral crystal structure, *R3c*. Furthermore, we determined the direct band-gap energy ($E_g$), the separation between the valence band and Fermi energy ($E_V$-$E_F$), and Urbach energy ($E_U$) for the BiFeO$_{2.85}$ film to be approximately 2.27 eV, 0.9 eV, and 0.38 eV, respectively. Based on these parameters, we proposed an electronic energy-level diagram for the BiFeO$_{2.85}$ film. Additionally, we observed weak ferromagnetic behavior for the BiFeO$_3$ film within the temperature range of 2–300 K. This behavior is attributed to the distortion of FeO$_6$ octahedra in the film due to the induced oxygen vacancies, which leads to the disruption of a short-range spin cycloid structure in rhombohedral BiFeO$_{2.85}$ thick film.

## 2. EXPERIMENTAL DETAILS

First, we prepared a Bi-rich Bi$_{1.1}$FeO$_3$ pellet in disc form (12 mm in diameter and 2 mm in thickness) using a conventional solid-state reaction. Subsequently, BiFeO$_3$ films were grown on an *n*-type Si(100) substrate at various substrate temperatures ($T_S$) and oxygen partial pressures ($O_{2pp}$) using the pulsed laser deposition (PLD) technique.[18,19] A Nd: YAG-based pulsed laser with a fluence of 2.7 J/cm$^2$, pulse width of 19 ns, wavelength of 355 nm, and frequency repetition of 10 Hz was used to grow the BiFeO$_3$ films on Si(100) substrates. The target and substrates were fixed in their respective positions in the vacuum chamber, which was then evacuated to a base pressure of 5 × 10$^{-6}$ mbar using rotary and diffusion pumps. The substrates were heated to temperatures ranging from $T_S$ = 525 ºC to $T_S$ = 625 ºC with 25 ºC increments using a heater attached to the substrate holder. Oxygen gas was partially filled in the chamber to reach a constant $O_{2pp}$ of 0.06 mbar during the growth of BiFeO$_3$ films at the different substrate temperatures. The growth time for all films was 10 minutes. After growth, the films were in-situ annealed at the respective growth temperature with $O_{2pp}$ of 100 mbar for 1 hour, followed by cooling to room temperature under the same $O_{2pp}$. The X-ray diffraction (XRD) patterns of the grown films were recorded (with a scan rate of 0.016º) and analyzed using a PANalytical X'Pert X-ray diffractometer with Cu $K\alpha$ ($\lambda$ = 1.5406 Å) X-ray line. The results based on the XRD analysis of all grown films are given in section 3.1. From this analysis, it was apparent that the optimized growth condition was $T_S$ = 575 ºC with $O_{2pp}$ = 0.06 mbar to grow a single-phase BiFeO$_3$ film.

The microstructure of the optimized BiFeO$_3$ film grown at $T_S$ = 575 ºC and $O_{2pp}$ = 0.06 mbar was examined using scanning electron microscopy (SEM) and atomic force microscopy (AFM) at room temperature. Subsequently, the film was subjected to temperature- and magnetic field ($H$)-dependent magnetization ($M$) measurements ($M$ vs. $T$ and $M$ vs. $H$) using a physical property measurement system (PPMS from Quantum Design, USA). Following this, the Raman spectrum of the BiFeO$_3$ film was obtained at room temperature in a backscattering geometry using a Horiba-Jobin-Yvon HR8000 Raman spectrometer with a He-Ne laser emitting at a wavelength of 632.8 nm. Lastly, a comprehensive analysis of the optical and elemental properties of the BiFeO$_3$ film at room temperature was conducted using a Cary 5E UV-visible-NIR spectrophotometer and a Perkin Elmer X-ray photoelectron spectrometer (XPS), respectively.

## 3. RESULTS AND DISCUSSION

*3.1 Structural characterization*

The XRD patterns of all grown films in this study are available in the supplementary information (refer to Figs. S1 and S2). A semi-log plot of a phase diagram illustrating the relationship between oxygen pressure and growth temperature was plotted to show the formation of different phases, including the desired rhombohedral phase of $BiFeO_3$ (left inset of Fig. 1). It was observed that the secondary phase $Bi_4Fe_7O_9$ was mainly formed along with the desired $BiFeO_3$ phase at a low temperature of $T_S = 525$ °C. Conversely, the film grown at a high temperature of $T_S = 625$ °C was largely amorphous, although the $BiFeO_3$ phase was still present. The preferred phase of $BiFeO_3$ was found to be primarily formed at intermediate temperatures, $T_S = 550–600$ °C (Fig. S1). The film grown at $T_S = 575$ °C exhibited better crystalline nature compared to the films grown at other intermediate temperatures, as indicated by the XRD study (left inset of Fig. 1). Subsequently, $BiFeO_3$ films were deposited at different $O_{2pp}$ ($6 \times 10^{-3}$–3 mbar) while maintaining a fixed temperature of $T_S = 575$ °C (Fig. S2). The film grown at a high $O_2$ pressure of 3 mbar showed the presence of parasitic phases such as $Bi_2O_3$ and $Fe_2O_3$, while the film grown at a low $O_2$ pressure of $6 \times 10^{-3}$ mbar was predominantly amorphous with low-intensity peaks of $BiFeO_3$. The XRD analyses of these films revealed that the film grown at $T_S = 575$ °C with $O_{2pp} = 0.06$ mbar was optimized for growth conditions to obtain a single-phase rhombohedral $BiFeO_3$ film (see Figs. S1, S2, and 1). These observations align with a report by Bea et al.[18]

We estimated the percentage of desired $R3c$ $BiFeO_3$ and secondary phases in the films grown at different temperatures with $O_{2pp} = 0.06$ mbar, as presented in Table 1. It was found that the percentage of desired $R3c$ $BiFeO_3$ gradually increased from 50.7% to 100% until the temperature $T_S = 575$ °C, while the secondary phase $Bi_2Fe_4O_9$ was formed at low substrate temperatures, $T_S \leq 550$ °C. However, at high substrate temperatures, $T_S > 575$ °C, the parasitic phase $Bi_2O_3$ starts to crystallize along with the prepared *R3c* BiFeO$_3$. These findings are consistent with the results of Bea *et al*.[18] We also determined the out-of-plane lattice constant and microstrain of these films for comparison, which are listed in Table 1. We observed that the out-of-plane lattice parameter tends to slowly increase from 3.898 Å to 3.946 Å with increasing $T_S$. These obtained out-of-plane lattice parameters of the BiFeO$_3$ films are consistent with the literature.[14] While the induced microstrains in the films are evaluated using the Williamson-Hall equation,[20] $\varepsilon = \frac{\beta_{hkl}}{4\tan\theta}$, which are in the range of 0.15% to 0.37%. These observations are similar to the results of a BiFeO$_3$ thick film by Zhu *et al*.[17] Additionally, we noted that films grown at low temperatures $T_S \leq 575$ °C exhibit the preferred growth direction along the plane (110) through the estimated intensity ratio, $\frac{I_{(110)}}{I_{(100)}}$ (see Table 1), while the films grown at $T_S > 575$ °C have preferential growth along the plane (100). This finding will be further investigated by evaluating the degree of orientation of the optimized BiFeO$_3$ film grown at $T_S = 575$ °C and $O_{2pp} = 0.06$ mbar in the following.

The XRD pattern of the BiFeO$_3$ film grown on Si (100) at $T_S = 575$ °C and $O_{2pp} = 0.06$ mbar is shown in Fig. 1. It is evident that the film is polycrystalline with a preferred orientation along the (110) plane. All XRD peaks observed in this film can be indexed to a rhombohedral (*R*-distorted perovskite phase) crystal structure (PDF#73-0548). The values of lattice constants *a* and *c* of our film are determined to be 5.31 Å and 14.16 Å, respectively. To quantify the (110) orientation, the degree of orientation ($\alpha_{110}$) is calculated using the formula:[14]

$$\alpha_{110} = \frac{I(110) + I(210)}{\sum I(hkl)}. \qquad (1)$$

Where *I(hkl)* represents the total intensity of all observed XRD peaks of the *(hkl)* planes of BiFeO$_{2.85}$. The numerator in the formula includes the intensities of both parallel planes of (110) and (210). The calculated degree of orientation (*α$_{110}$*) of the BiFeO$_{2.85}$ film is approximately 52.9%.

The right inset in Fig. 1 shows the SEM image of the BiFeO$_{2.85}$ film. It reveals that the film's surface had uniform and densely packed grains with an average grain size of approximately 300 nm. By analyzing the energy-dispersive X-ray spectrum (EDS; Fig. S3), we determined the Bi/Fe ratio of the film to be approximately 1:1. Similarly, the oxygen content was found to be around 2.85, leading us to identify the composition of the film as BiFeO$_{2.85}$. This observation indicates oxygen deficiencies in the PLD-grown BiFeO$_3$ film, which will be further examined using the core-level *O 1s* XPS spectrum (see Fig. 8). To gain a better understanding of the film's microstructure, we captured 2D and 3D images (Fig. 2) of the film using the AFM microscope. This analysis confirms the preferred growth nature of the film observed in the XRD data. The average grain size of the BiFeO$_{2.85}$ film is estimated to be approximately 320 nm, consistent with the SEM study. Additionally, the surface roughness $R_{RMS}$ of the film is measured to be around 18 nm.

*3.2 Magnetization measurements*

Figure 3 illustrates the zero field-cooled (ZFC) and field-cooled warming (FC) magnetization of the oxygen-deficient BiFeO$_{2.85}$ film measured at an applied magnetic field of 100 *Oe* from 2 K to 300 K. It is observed that the ZFC and FC magnetization curves of the film split below 253 K with a growing divergence between them as the temperature decreases. The FC magnetization shows a steady increase with temperature until 5 K, followed by a sharp rise in magnetization below that temperature. On the other hand, the ZFC magnetization decreases gradually with decreasing temperature until 2 K. These observations suggest weak ferromagnetic behavior of the BiFeO$_{2.85}$ film below room temperature.[18,21,22] This will be further investigated using magnetic isotherms (*M* vs. *H*) curves at temperatures of 2, 50, 100, 200, and 300 K, which are presented in Fig. 4. All measured magnetic isotherms display weak ferromagnetic characteristics, likely due to strain-induced modulation of the magnetic structure of the $BiFeO_{2.85}$ film, as the induced microstrain in the thick film is found to be about 0.31% (see Table 1). For this film, a maximum saturation magnetization ($M_s$) and coercive field ($H_c$) of approximately 0.07 $\mu_B/f.u.$ and 206.7 $Oe$, respectively were observed at 2 K, consistent with previous studies on $BiFeO_3$ films.[18,21,22] These observations on the $BiFeO_3$ thick film are also similar to its bulk counterpart.[3,23-25]

Notably, both $M_s$ and $H_c$ values of the $BiFeO_{2.85}$ film are found to increase with decreasing temperature. Figure 5 displays the temperature-dependent coercive field ($H_c$) and exchange bias field ($H_{EB}$). These values were determined using the relations: $H_c = \frac{1}{2}(H_{c1} + H_{c2})$ and $H_{EB} = \frac{1}{2}(H_{c1} - H_{c2})$.[26] We observed an increase in $H_c$ with decreasing temperature for the oxygen-deficient $BiFeO_{2.85}$ film (Fig. 5 a), which is consistent with the FC magnetization data in Fig. 3. While the exchange bias field slightly increased to $H_{EB}$ = +6.45 $Oe$ at 200 K compared to the value ($H_{EB}$ = +4.67 $Oe$) at 300 K (Fig. 5b). Below 200 K, the $H_{EB}$ value was found to change to a negative bias field (-$H_{EB}$ = -8.12 $Oe$) at 100 K from the positive field (+$H_{EB}$) at $T \geq 200$ K and then it increases gradually with further decreasing temperature. The switching of $H_{EB}$ from positive to negative below 200 K is essentially due to the microstrain-induced modification in the magnetic structure of the oxygen-deficient $BiFeO_{2.85}$ film. These observations indicate that the spin cycloid in our thick film was only nearly destroyed by the induced microstrain (0.31%) in the (110)-textured oxygen-deficient $BiFeO_{2.85}$ film. Since the destruction of the short-range magnetic ordering in $BiFeO_3$ requires a slightly higher microstrain > 0.5% according to Sando *et al*.[27]

To further analyze the magnetic behavior of the oxygen-deficient $BiFeO_{2.85}$ film, we plotted the Arrott-Belov-Kouvel (ABK) plots[28-30] ($M^2$ vs. $H/M$, as presented in Fig. 6) based on Weiss molecular field theory ($\frac{H}{M} = 2a + 4bM^2$) for all measured temperatures using the magnetic isotherms (see Fig. 4). Here, factors *a* and *b* are temperature-dependent coefficients. We observed that the ABK plots of the *M(H)* data at different temperatures exhibit a convex feature with positive slopes.[29] Additionally, our oxygen-deficient BiFeO$_{2.85}$ thick film has a non-zero spontaneous magnetization at *H* = 0 Oe for all measured temperatures (see inset of Fig. 4). These findings confirm a weak ferromagnetic behavior of the film in the temperature range of 2–300 K (see Figs. 3–5),[28-30] however both ferromagnetic and antiferromagnetic orders coexist and compete with each other in the BiFeO$_{2.85}$ film. Therefore, the negative bias field was detected for the inspected film at *T* < 200 K, while the positive field was observed at *T* ≥ 200 K (see Fig. 5b).

*3.3 Raman spectroscopy study*

Figure 7a displays the Raman spectrum of the BiFeO$_{2.85}$ film on Si (100) recorded at room temperature in the wavenumber range of 100–600 cm$^{-1}$. The observed high-intensity Raman mode at 523.5 cm$^{-1}$ is attributed to the Si (100) substrate.[12] Among the 13 (9*E*+4*A$_1$*) Raman-active modes allowed by the *R3c*-symmetry of BiFeO$_3$,[12,31-34] we observed 11 allowed Raman modes (7 *E(TO)* and 4 *A$_1$(TO)*). Specifically, the modes near 150, 220, 310, and 555 cm$^{-1}$ belong to *A$_1$(TO2)*, *A$_1$(TO2)*, *A$_1$(TO3)*, and *A$_1$(TO4)*, respectively. The modes detected near 130, 240, 260, 280, 345, 380, and 440 cm$^{-1}$ correspond to the seven allowed Raman modes such as *E(TO2)*, *E(TO3)*, *E(TO4)*, *E(TO5)*, *E(TO6)*, *E(TO7)*, and *E(TO8)*, respectively. Additionally, the four *LO* modes, namely *A$_1$(LO1)*, *A$_1$(LO4)*, *E(LO2)*, and *E(LO8)* modes were observed due to the splitting of transverse and longitudinal phonons (TO-LO) related to the 4 Raman-active modes of the *R3c*-symmetry of BiFeO$_3$.[31-34] We did not detect the first *E(TO1)* that usually appear below 100 cm$^{-1}$,[31] mainly due to the limitations of our Raman spectrometer.[33] Besides, the last *E(TO9)* mode (which usually occurs near 523 cm$^{-1}$)$^{35}$ is not detected due to the high-intensity Raman mode of the Si(001) substrate at 523.5 cm$^{-1}$.

To evaluate the frequency of Raman modes in the BiFeO$_{2.85}$ film, we deconvoluted the Raman spectrum of the film below 400 cm$^{-1}$ using the PeakFit software package, as displayed in Fig. 7b. The frequencies of six *E(TO)* modes are approximately 135.5, 239.4, 263.4, 285.3, 345.8, and 380.3 cm$^{-1}$ corresponding to *E(TO2)*, *E(TO3)*, *E(TO4)*, *E(TO5)*, *E(TO6)*, and *E(TO7)*, respectively. The frequencies of three *A$_1$(TO)* modes are 156.5, 218.5, and 307.3 cm$^{-1}$ which are related to *A$_1$(TO2)*, *A$_1$(TO2)*, and *A$_1$(TO3)* modes, respectively. Additionally, the frequencies of four *LO* modes, *A$_1$(LO1)*, *E(LO2)*, *E(LO5)*, and *E(LO6)* due to the *LO-TO* splitting were detected at approximately 174.3, 187.7, 328.9, and 363.5 cm$^{-1}$. These findings are in excellent agreement with the results of Hlinka *et al*.[35] Here, a goodness-of-fit with the coefficient of determination, $R^2$ =0.998 was achieved by including two additional phonon modes at 120.6 and 199.5 cm$^{-1}$. The former mode is usually detected in BiFeO$_3$ polycrystals,[34] but not in single crystals or thin films.[31-35] The latter mode at 199.5 cm$^{-1}$ may be related to an extraordinary mode that occurs due to the mixing between *E(TO)* and *A$_1$(TO)* modes according to the observations of Talkenberger *et al*.[34] This finding needs further investigation to conclude.

*3.4 UV-visible-NIR spectroscopy study*

To investigate the optical properties of the grown BiFeO$_{2.85}$ film, we used the same growth conditions to coat the film on a quartz substrate (refer to Section 2). This allowed us to measure the reflectance (*R*) and transmittance (*T*) of the film simultaneously as a function of wavelength (*λ*). By analyzing the reflectance spectrum, *R* versus *λ* of the film within the wavelength range of 500–3300 nm (the top inset of Fig. 8), we estimated the thickness (*d*) of the BiFeO$_3$ film using an interference interval method with a refractive index of *n* = 2.5 to be approximately 2 μm based on the observed fringe patterns in the reflectance spectrum. This finding indicates that our film is relatively thick. We observed a rapid decrease in the reflectance below 800 nm, suggesting strong absorption in the range of 400–800 nm, covering the entire visible spectrum. This indicates the potential of our film for optoelectronic applications.[36,37]

Furthermore, by analyzing the measured $R(\lambda)$ and $T(\lambda)$ spectra (refer to the top inset of Fig. 8 and Fig. S4), we evaluated the absorption coefficient ($\alpha$) as a function of wavelength using the absorbance ($A$) relation with $R$ and $T$,[36]

$$A = \alpha d = -\ln\frac{(1-R)^2}{T}. \quad (2)$$

Subsequently, we plotted $(\alpha E)^2$ (left y-axis) and $(\alpha E)^{1/2}$ (right y-axis) as a function of incident photon energy ($E=h\nu$) for the BiFeO$_{2.85}$ film, as shown in Fig. 8. These plots were created to determine if the $\alpha E$ data of the BiFeO$_{2.85}$ thick film obeys the relation:[36-40]

$$\alpha h\nu = A(h\nu - E_g)^n. \quad (3)$$

Where $E_g$ is the optical band-gap energy, $A$ is a constant, and parameter $n$ is a power factor indicating whether the band-gap transition is direct ($n = 1/2$) or indirect ($n = 2$). The estimated Tauc gap (onset of optical absorption) for the inspected film is approximately 2.18 eV, which aligns with the values (2.11-2.15 eV) reported for the epitaxial BiFeO$_3$ films.[40,41] Furthermore, the direct band gap of the BiFeO$_{2.85}$ thick film is determined to be $E_g$=2.27 eV through the linear extrapolation of the $(\alpha E)^2$ value tending to zero. This value is comparable to the reported value ($E_g$=2.32 eV) for the BiFeO$_3$ film.[42] However, it is much lower when compared with the values ($E_g$=2.5–2.8 eV) of the high-crystalline BiFeO$_3$ bulk and thin films.[5,40-42] Conversely, the indirect band gap of the thick film is calculated to be about 1.93 eV. This is likely attributed to the influence of the electronic structure features in the film due to the induced defects such as oxygen vacancies, grain boundaries, and strain.[23,39-43]

To verify this observation, we plotted the graph of ln$\alpha$ versus $E$ (bottom inset of Fig. 8) to calculate the Urbach energy ($E_U$) using the Urbach empirical formula:[39]

$$\alpha = \alpha_0 e^{\frac{E}{E_U}}. \qquad (4)$$

Here, $\alpha_0$ is a constant. Estimating the Urbach energy provides information about the defects associated with the intermediate energy levels close to the conduction band minimum (CBM) and/or valence band maximum (VBM) of $BiFeO_3$.[39] Our evaluation shows that the $E_U$ value is approximately 0.38 eV for the $BiFeO_{2.85}$ thick film, consistent with the energy level of shallow oxygen vacancy states (0.3 eV) rather than deep trap levels ($\geq$ 0.6 eV) due to the presence of $Fe^{2+}$ ions in $BiFeO_3$.[44-46] Importantly, this $E_U$ value is similar to the energy difference (0.34 eV) between the direct and indirect band gap energies of our film. Therefore, the observed indirect band transition is mainly due to the presence of oxygen vacancies and other defects in our $BiFeO_{2.85}$ film. This will be further analyzed using the core-level XPS spectra of the Fe 2*p* and *O* 1*s* peaks of the $BiFeO_{2.85}$ film along with its valence band spectrum (refer to section 3.5).

*3.5 X-ray photoelectron spectroscopy study*

The XPS survey scan of the $BiFeO_{2.85}$ thick film in the energy range of 0–1050 eV is presented in Fig. 9. The XPS survey spectrum of the film was recorded using an Al *K$\alpha$* x-ray line (with 1486 eV) at room temperature, using a concentric hemispherical energy analyzer with pass energy of 50 eV. This measurement was conducted after cleaning the film's surface with a 3 keV argon ion ($Ar^+$) gun sputtering to remove surface contamination as the film was exposed to air during handling for characterization. More details about the XPS measurements were reported elsewhere.[46] Additionally, the binding energies of the core-level XPS peaks of the constituent ions, $Bi^{3+}$, $Fe^{3+}$, and $O^{2-}$ of the sample were calculated after calibrating their spectra using a standard of the surface carbon feature C *1s* at 284.6 eV (see Fig. S5).

Here, the characteristic peaks of Bi $4f_{7/2}$ (157.8 eV), Bi $4f_{5/2}$ (163.2 eV), Fe $2p_{3/2}$ (719.8 eV), Fe $2p_{1/2}$ (723.6 eV), and O $1s$ (529.6 eV) from the constituent ions were detected in the BiFeO$_{2.85}$ film. Additionally, the features of Bi $5d$ (25.6 eV), Fe $3p$ (55.6 eV), Fe $3s$ (92.6 eV), Bi $5p$ (118.6 eV), Bi $4d$ (440.6 & 464.6 eV), and Bi $4p$ (679.6 & 804.6 eV) were observed for our film. Notably, the energy difference ($\Delta E$) between Bi $4f_{7/2}$ and Bi $4f_{5/2}$ is found to be about 5.4 eV (See inset of Fig.8). These findings are consistent with literature data.[47-49] Furthermore, X-ray-induced Auger electron features; particularly Fe $L_3M_{45}M_{45}$, Fe $L_3M_{23}M_{45}$, Fe $LM_{23}M_{23}$, and O $KLL$ were visibly noticed near 781.6, 845.6, 898.6, and 972.6 eV, respectively.

Figure 10 displays the fitted narrow-scan O $1s$ XPS spectrum of the probed film after Shirley-type background subtraction. A good fit is achieved using three components at 529.6, 530.9, and 532.1 eV. The first two components are attributed to the metal-oxygen (M-O$_x$) bonds, specifically the Fe-O and Bi-O bonds, which align with our previous results on bulk BiFeO$_3$.[47] The third component, at 532.1 eV, is typically linked to oxygen vacancy defects caused by dangling bonds in the BiFeO$_3$ material.[49] Notably, the contribution of the Bi-O bond (37.5%) to the O $1s$ peak is lower compared to the contribution of the Fe-O bond (46.8%), likely due to Ar+-ion sputtering-induced metal Bi into the film (see Fig. S6). This results in the dominant behavior of the Fe-O bond over the Bi-O bond; in line with our findings on single-phase bulk BiFeO$_3$.[47] Our estimation indicates that the oxygen vacancies in the film are approximately 15.7%. Overall, the ratio between lattice oxygen ($O_L$) and oxygen vacancy ($O_V$) is estimated to be approximately 0.19 for our film. These observations are consistent with our EDS analysis.

The recorded narrow-scan Fe $2p$ spectrum of the BiFeO$_{2.85}$ thick film in the range of 705–740 eV is presented in Fig. 11a. Two features belonging to the Fe $2p_{3/2}$ and Fe $2p_{1/2}$ were observed near 710 and 724 eV, respectively. Interestingly, the $\Delta E$ value between the Fe $2p_{3/2}$ and Fe $2p_{1/2}$ peaks is about 14 eV, indicating that the Fe ions in the film are $Fe^{3+}$ ions.[47,50] Additionally, two satellite peaks related to these two characteristic peaks were detected near 718 and 732 eV, respectively. However, no distinct feature near 709 eV, which is usually related to $Fe^{2+}$ $2p_{3/2}$ was observed for our sample. Therefore, we conclude that the Fe ions in the inspected film are present as $Fe^{3+}$ valence states. The fitting was performed using symmetrical or asymmetrical Gaussian-Lorentzian (GL) product function with the help of the RBD's AugarScan 3.2 software, which is described in detail in our earlier work.[47] This fitting helped us deduce the characteristic line widths and shapes of the Fe $2p$ core-level spectrum for the sample.

Figure 11b displays the GL function fitted curve of the Shirley-type background subtracted $Fe^{3+}$ $2p_{3/2}$ peak of the thick film. This non-linear fit included seven features: a pre-peak, four-allowed Gupta and Sen (GS) multiplets of $Fe^{3+}$ $2p_{3/2}$, a surface peak, and a satellite peak of $Fe^{3+}$ $2p_{3/2}$. The individual curves of the resultant fit and seven sub-peaks, along with the measured data are shown in Fig. 11b. The obtained binding energy (BE) and line width ($w$) of the sub-peaks of the $BiFeO_{2.85}$ thick film, along with their $\Delta E$ values, are presented in Table 2. For comparison, the data of the compounds (from Refs. 47 and 50), namely bulk $BiFeO_3$, $Bi_{0.9}Ba_{0.1}FeO_{2.95}$, $Bi_{0.9}Ba_{0.05}Ca_{0.05}FeO_{2.95}$, and $\gamma$-$Fe_2O_3$, are also listed in Table 2. The estimated BE and $w$ values of four GS multiplets, along with other sub-peaks of our film, are in good agreement with the literature.[47,50]

In particular, the $w$ values (1.3–1.5) of our film are comparable to those of bulk $BiFeO_3$ materials (1.2–1.7).[47] Furthermore, the combined line width (8.4 eV) of the four GS multiplets of $Fe^{3+}$ $2p_{3/2}$ peaks in the presently inspected non-stoichiometric $BiFeO_{2.85}$ film is lower (9.5 eV) when compared with its bulk stoichiometric counter-part, $BiFeO_3$.[47] However, it is comparable to that of the oxygen-deficient bulk $Bi_{0.9}Ba_{0.1}FeO_{2.95}$ (8.8 eV).[47] Additionally, the intensity (116) of its satellite feature is also similar (111) to the bulk $Bi_{0.9}Ba_{0.1}FeO_{2.95}$. Moreover, we found that the $\Delta E$ value between the $Fe^{3+}$ $2p_{3/2}$ peak of our film and its satellite feature is slightly lower (7.8 eV) when compared with the stoichiometric $BiFeO_3$ ceramic (8.1 eV) and $\gamma$-$Fe_2O_3$ (8.3 eV).[47,50] However, this value is comparable to the value (7.7 eV) of the non-stoichiometric compound, $Bi_{0.9}Ba_{0.05}Ca_{0.05}FeO_{2.95}$.[47] These observations validate the oxygen deficiency detected in the PLD-grown $BiFeO_3$ thick film. Thus, we evaluated the average ligand electronegativity of oxygen (O) to be about 3.26, which is smaller than that of other listed materials (see Table 2).[47,50] We included the high-BE surface peak and low-BE pre-peak in the non-linear fitting of the $Fe^{3+}$ $2p_{3/2}$ spectrum of the studied film to obtain reliable physical parameters. These low-intensity high- and low-BE peaks in the core-level Fe 2p spectrum usually occur due to the $Ar^+$-ion sputtering of the surface of the $BiFeO_3$.[47]

The experimental valence band spectrum of the $BiFeO_{2.85}$ film is shown in Fig. 12. The broad peak observed at 5.4 eV is primarily due to the $t_{2g}$ and $e_g$ states resulting from a strong hybridization of the Fe 3d-O 2p states, along with contributions from Bi 6p states.[51,52] These contributions are from both p-d and d-d charge transfer transitions in the $FeO_6$ octahedra of $BiFeO_3$. This finding is in excellent agreement with the findings of R-phase $BiFeO_3$.[47,51-53] It was observed that the valence band maximum, $E_V$ of the film is situated 0.9±0.1 eV below the Fermi energy $E_F$, which is lower than that of the stoichiometric $BiFeO_3$ thin film (1.2 eV).[54] This is essentially due to the presence of oxygen vacancies in our film (refer to Fig. 10), which introduce defect energy level at $E_U = 0.38$ eV below $E_C$ and/or above $E_V$. This disorder energy level (Urbach tail) may exist as a narrow electronic band within the band gap of the $BiFeO_3$.[55]

*3.6 An electronic band structure of the PLD-grown oxygen-deficient $BiFeO_{2.85}$ thick film*

Based on the current investigation, an energy level diagram is proposed and presented in Fig. 13 for the PLD-grown oxygen-deficient $BiFeO_{2.85}$ thick film based on the separation energy ($E_V$-$E_F$), band-gap energy ($E_g$), Tauc gap, and Urbach energy ($E_U$) obtained from experimental absorption and valence band spectra (see Figs. 8 and 12). It was observed that all core level spectra of the constituents, Bi, Fe, and O were affected by the induced oxygen vacancies, likely introducing acceptor levels above the VBM. Therefore, significant alterations in binding energy, full width at half maximum (FWHM, $w$), and the separation between spin-orbit splitting peaks in the Fe 2p spectrum were observed for the oxygen-deficient $BiFeO_{2.85}$ film when compared to bulk $BiFeO_3$ (refer to Table 2). Particularly, we attribute the Fermi energy shift towards the VBM of the $BiFeO_{2.85}$ film to the electronic shifts rather than chemical shifts (which only lead to a change in the oxidation state and binding energy of the elements).[56] Hence, band-gap energy, the separation ($E_V$-$E_F$) energy, and all core-level spectra of the constituents were considerably altered due to electronic shifts induced by oxygen vacancies in the $BiFeO_{2.85}$ film (refer to Figs. 8–12 and Table 2). As a result, the combined line width (8.4 eV) of four GS $Fe^{3+}$ $2p_{3/2}$ peaks and the separation energy ($\Delta E$) between the $Fe^{3+}$ $2p_{3/2}$ peak and its satellite peak were comparable to oxygen-deficient Ba- and (Ba, Ca)-substituted oxygen-deficient $BiFeO_3$ polycrystalline materials (see Table 2).[47] Consequently, the oxygen vacancies-induced global electronic shift of the core-level of the constituent atoms in the $BiFeO_{2.85}$ thick film was detected, which led to excellent absorption properties (i.e., weak reflectance behavior) in the range from 200 to 800 nm, making it a promising material for optoelectronic applications such as photovoltaic or photodetector devices.

## 4. CONCLUSION

The study investigated the structural, magnetic, and optical properties of an oxygen-deficient BiFeO$_{2.85}$ thick film grown on a Si (100) substrate using pulsed laser deposition under optimized growth conditions. The substrate temperature was set at $T_S$ = 575 ºC and the oxygen partial pressure at $O_{2pp}$ = 0.06 mbar to obtain a single-phase film. Structural analysis using XRD, EDS, and AFM studies revealed that the BiFeO$_{2.85}$ film grew as a rhombohedral (110)-textured film with an average grain size of 300 nm and a roughness of 18 nm. In the magnetic studies, the BiFeO$_{2.85}$ thick film shows weak ferromagnetic behavior between 2 and 300 K, with a non-zero spontaneous magnetization at $H$ = 0 Oe. This is most likely because the induced oxygen vacancies alter the magnetic structure of the BiFeO$_{2.85}$ film. With UV-visible-NIR spectroscopic analysis, the direct band-gap energy and Urbach energy for the examined film were calculated to be $E_g$ = 2.27 eV and $E_U$ = 0.38 eV, respectively. Oxygen vacancy defects were found in the recorded O 1$s$ XPS spectrum, which are primarily responsible for the created intermediate energy level, $E_U$. Core-level Fe 2$p$ XPS analysis revealed that the constituent Fe ions are present only in the Fe$^{3+}$ valence state. The Fermi energy is situated at 0.9 eV above the valence band maximum of the BiFeO$_{2.85}$ film, which red-shifted in comparison to its bulk counterpart because of the electronic shifts caused by the induced oxygen vacancies. An electronic energy-level diagram is finally suggested for the examined oxygen-deficient BiFeO$_{2.85}$ thick film.

 **Funding:**

This research did not receive any financial support from the funding agencies in the public, commercial, or not-for-profit sectors.

**CONFLICTS OF INTEREST:**

There are no conflicts of interest to declare.

**DATA AVAILABILITY STATEMENT:**

The data supporting this article have been included as part of the Supplementary Information.

**ORDCID**

**Ramachandran B** https://orcid.org/0000-0003-3676-1410

**Ambesh Dixit** https://orcid.org/0000-0003-2285-0754

**M.S. Ramachandra Rao** https://orcid.org/0000-0002-7806-2151

Table 1 the crystallographic parameters of the BiFeO$_3$ films grown at different temperatures with O$_{2pp}$ = 0.06 mbar.

| Growth temperature ($T_S$) | Desired phase $R3c$ BiFeO$_3$ | Secondary phases Bi$_2$Fe$_4$O$_9$ or Bi$_2$O$_3$ | $\frac{I_{(110)}}{I_{(100)}}$ | Out-of-plane lattice constant (Å) | Microstrain (%) |
|---|---|---|---|---|---|
| 525 ºC | 50.7% | 49.3% (Bi$_2$Fe$_4$O$_9$) | 0.61 | 3.898 | 0.37 |
| 550 ºC | 81.4% | 18.6% (Bi$_2$Fe$_4$O$_9$) | 0.76 | 3.936 | 0.25 |
| 575 ºC | 100 % | - | 1.46 | 3.939 | 0.31 |
| 600 ºC | 83.5% | 16.5 (Bi$_2$O$_3$) | 0.91 | 3.946 | 0.15 |
| 625 ºC | 50.7% | 49.3 (Bi$_2$O$_3$) | 1.06 | - | - |

Table 2 the obtained fitting parameters from the Fe $2p_{3/2}$ spectrum of the $BiFeO_3$-thick film with the factors of the bulk materials, namely $BiFeO_3$, $Bi_{0.9}Ba_{0.1}FeO_{2.95}$, $Bi_{0.9}Ba_{0.05}Ca_{0.05}FeO_{2.95}$, and $\gamma$-$Fe_2O_3$, for comparison.

| Peak/Sample | $BiFeO_{2.85}$-thick film (eV) [present work] | Bulk $BiFeO_3$ (eV) [Ref. 47] | Bulk $Bi_{0.9}Ba_{0.1}FeO_{2.95}$ (eV) [Ref. 47] | Bulk $Bi_{0.9}Ba_{0.05}Ca_{0.05}FeO_{2.95}$ (eV) [Ref. 47] | $\gamma$-$Fe_2O_3$ (eV) [Ref. 50] |
|---|---|---|---|---|---|
| Pre-peak [$w$] | 707.1 [2.1] | 708.0 [1.9] | 708.0 [1.9] | 707.4 [2.2] | - |
| $Fe^{3+}$ $2p_{3/2}$ peak 1 [$w$] | 709.8 [2.1] | 709.4 [2.3] | 709.6 [2.2] | 709.2 [2.5] | 709.8 [1.2] |
| $Fe^{3+}$ $2p_{3/2}$ peak 2 [$w$] | 711.3 [2.1] | 710.6 [2.4] | 710.5 [2.2] | 710.7 [2.4] | 710.8 [1.3] |
| $Fe^{3+}$ $2p_{3/2}$ peak 3 [$w$] | 712.6 [2.1] | 712.0 [2.4] | 712.0 [2.1] | 712.1 [2.6] | 711.8 [1.4] |
| $Fe^{3+}$ $2p_{3/2}$ peak 4 [$w$] | 713.9 [2.1] | 713.5 [2.4] | 713.6 [2.3] | 713.8 [2.3] | 713.0 [1.4] |
| Surface peak [$w$] | 715.7 [1.9] | 715.6 [2.1] | 715.7 [2.0] | 715.7 [1.9] | 715.0 |
| $Fe^{3+}$ satellite peak [$w$] | 717.6 [1.9] | 717.5 [2.0] | 717.6 [1.8] | 716.9 [2.0] | 718.1 |
| $I_{Satellite}$ of $Fe^{3+}$ [Area] | 116 [235] | 132 [282] | 111 [215] | 298 [628] | - |
| $\Delta E$(satellite-peak 1) | 7.8 | 8.1 | 8.0 | 7.7 | 8.3 |
| $\Delta E$(peak 2-peak 1) | 1.5 | 1.2 | 1.1 | 1.5 | 1.0 |
| $\Delta E$(peak 3-peak 2) | 1.4 | 1.4 | 1.5 | 1.4 | 1.0 |
| $\Delta E$(peak 4-peak 3) | 1.3 | 1.5 | 1.6 | 1.7 | 1.2 |
| Average ligand (O) electronegativity | 3.26 | 3.44 | 3.38 | 3.38 | 3.5 |

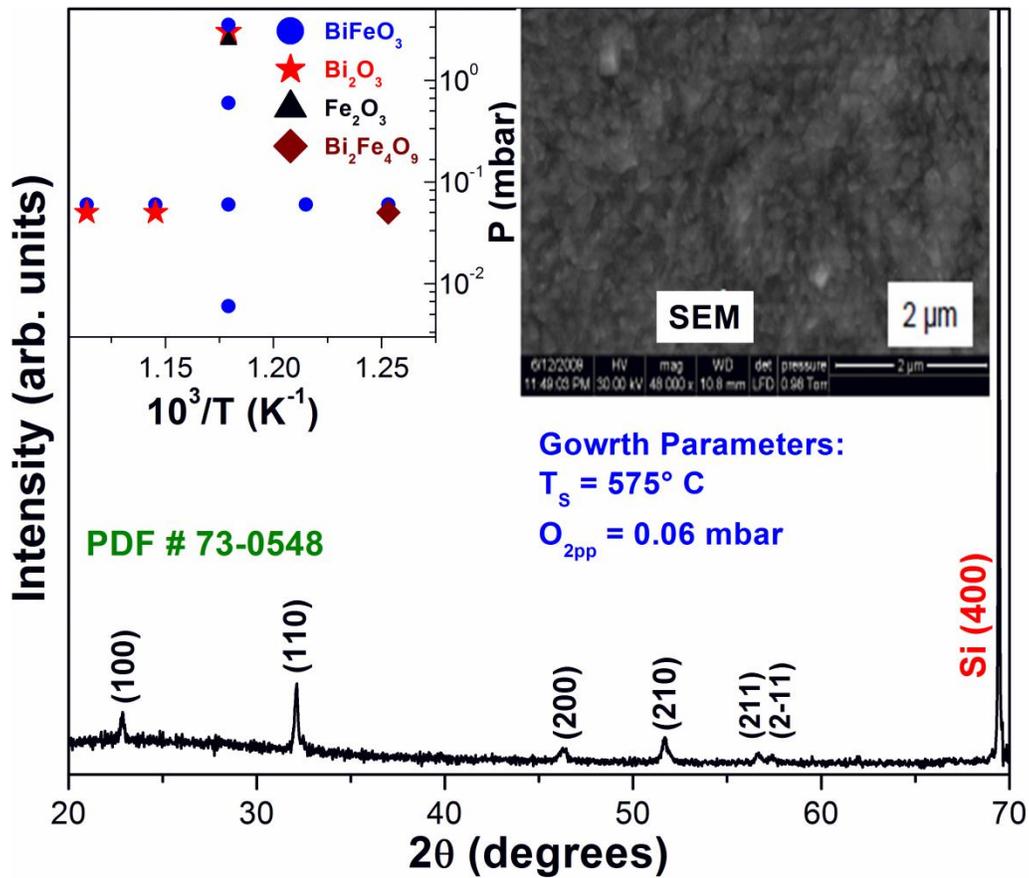

**Fig. 1** The X-ray diffraction pattern of the BiFeO$_3$ film grown at $T_S$ = 575 °C with $O_{2pp}$ = 0.06 mbar. The scanning electron microscope image is shown in the right inset, while the left inset displays a semi-log plot of the pressure-temperature phase diagram for the growth of the BiFeO$_3$ films.

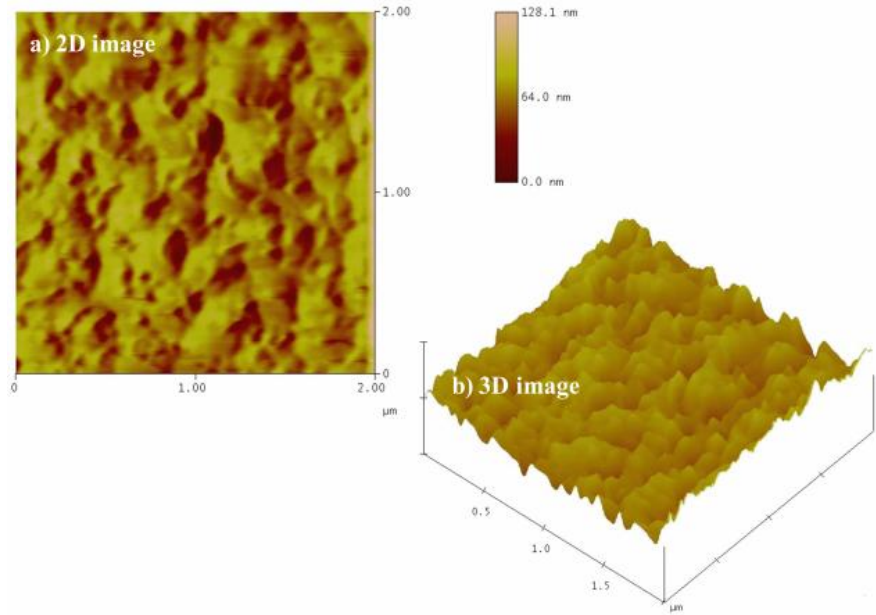

**Fig. 2** Atomic force microscopic images of the BiFeO$_{2.85}$ film in both a) 2D and b) 3D.

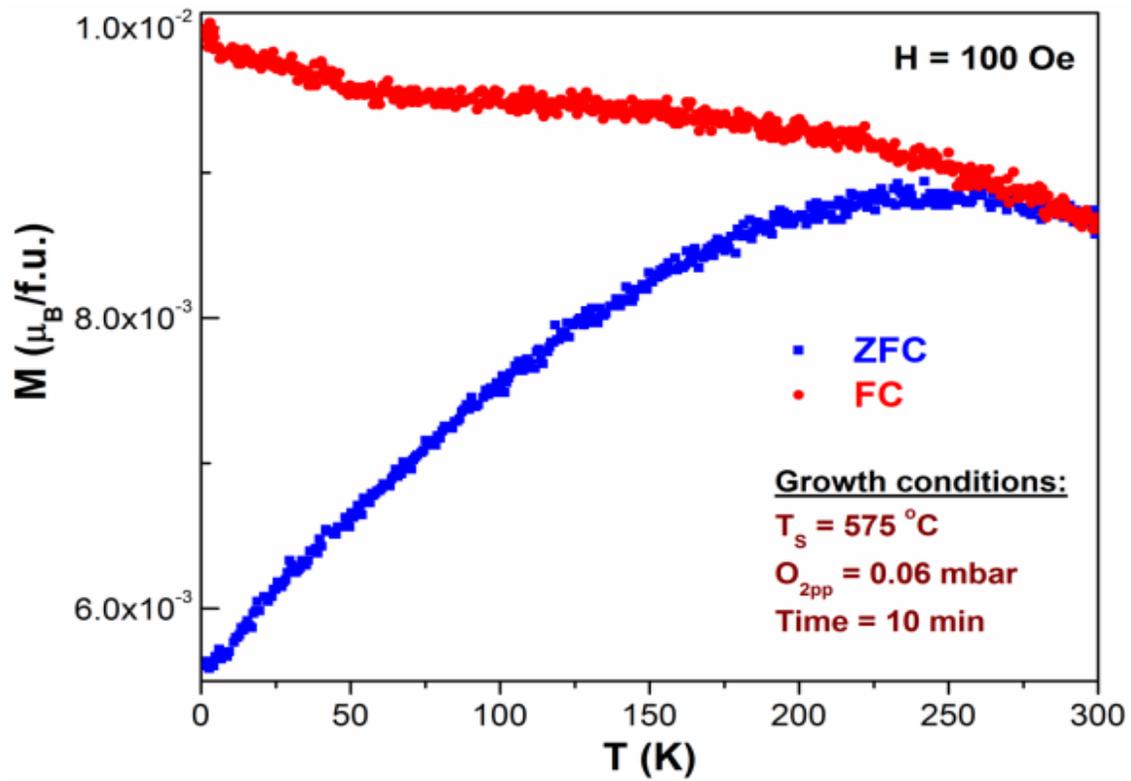

**Fig. 3** The temperature-dependent zero-field- and field-cooled (ZFC and FC) magnetization curves of the BiFeO$_{2.85}$ thick film measured at the field, $H = 100$ Oe in the temperature range of 2–300 K.

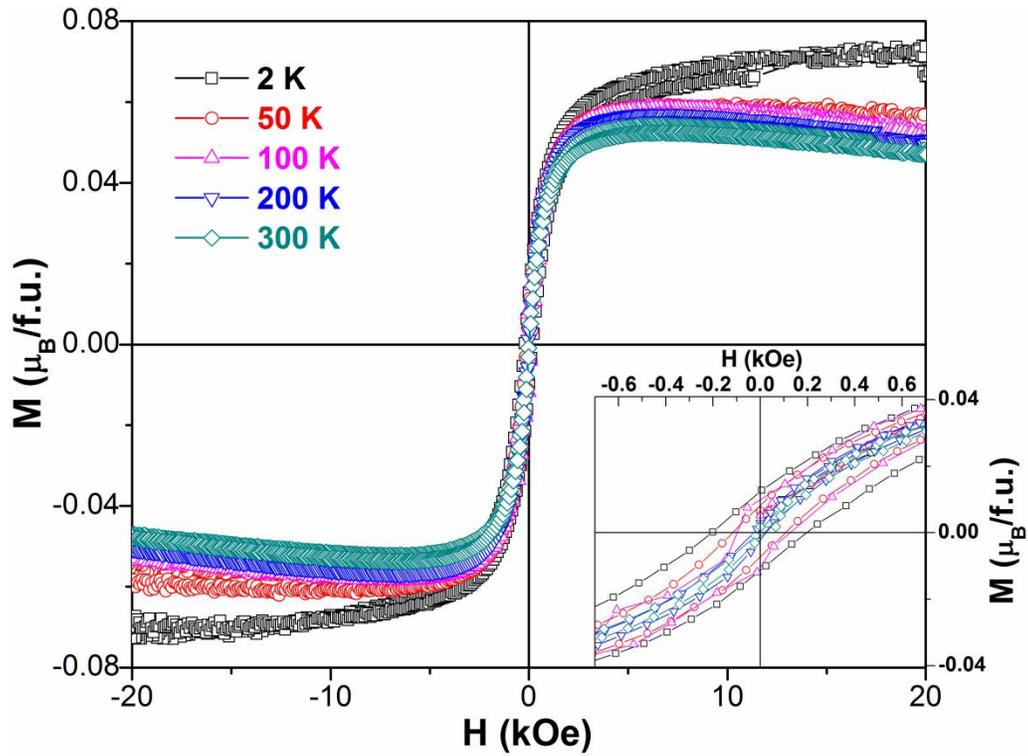

**Fig. 4** Magnetization (*M*) versus magnetic field (*H*) curves of the BiFeO$_{2.85}$ film recorded at different temperatures. The inset displays the corresponding low-field *M-H* curves.

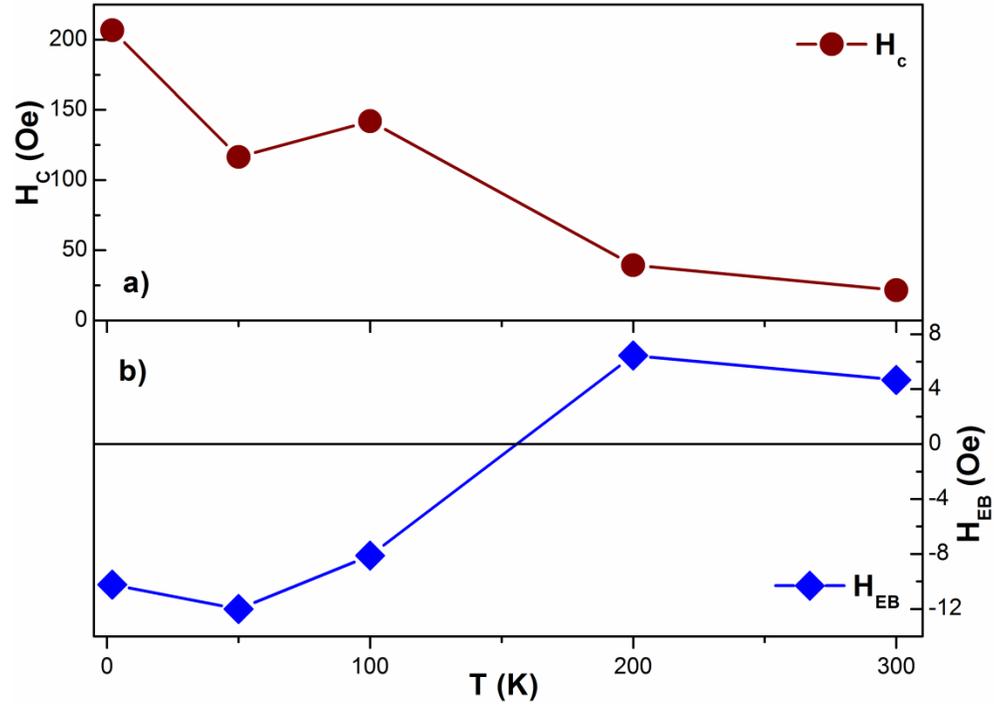

**Fig. 5** a) the coercivity ($H_C$) and b) exchange bias field ($H_{EB}$) versus temperature ($T$) for the BiFeO$_{2.85}$ thick film. The solid lines in the insets are provided for visual reference.

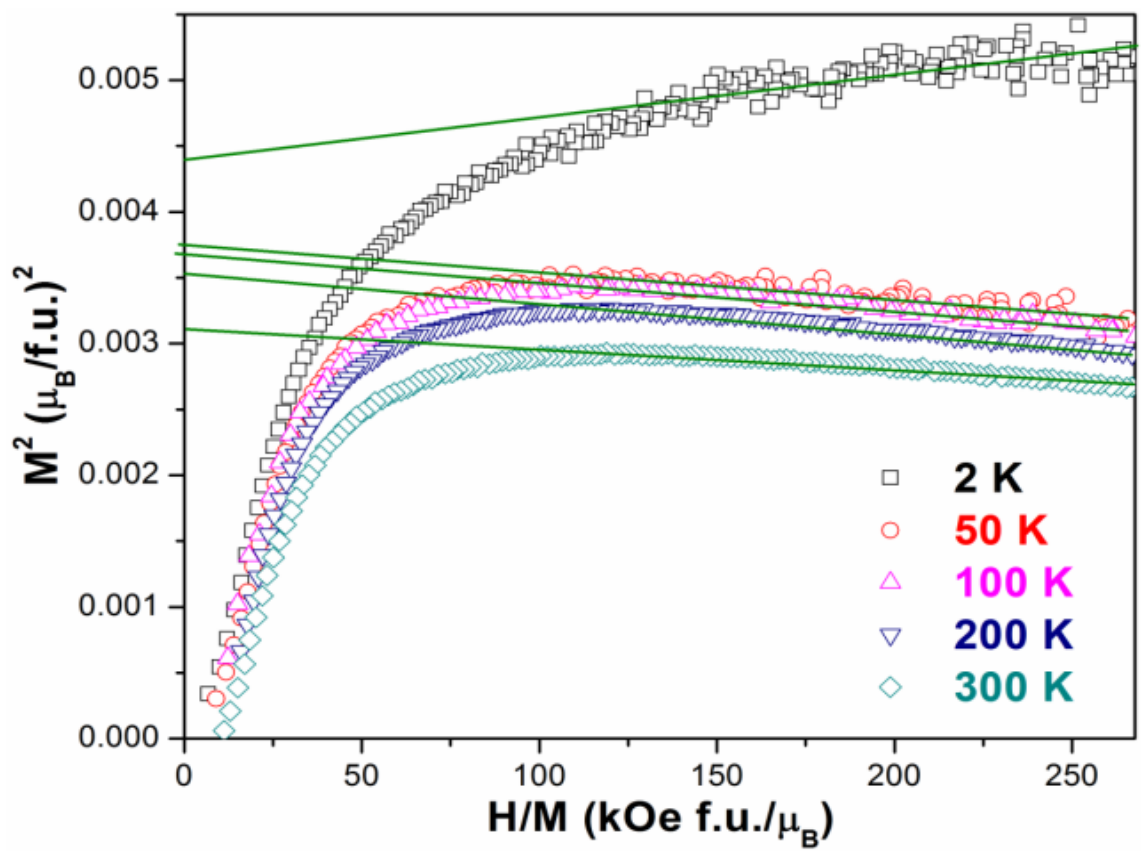

**Fig. 6** The Arrott-Belov-Kouvel (ABK) plot of the BiFeO$_{2.85}$ film using the magnetic isotherms at different temperatures is shown in Figure 4. The solid lines correspond to the linear fit to the respective ABK curves.

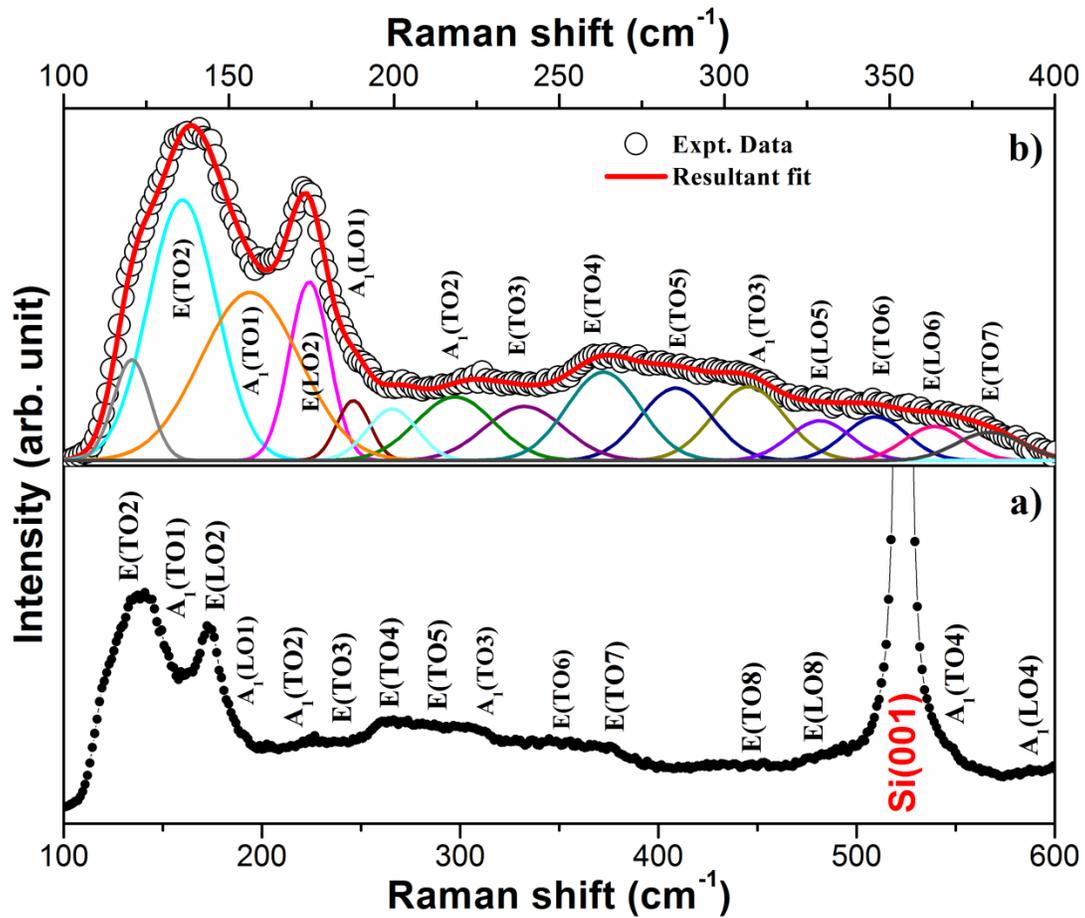

**Fig. 7** a) The Raman spectrum of the BiFeO$_{2.85}$ film in the wavenumber range of 100–600 cm$^{-1}$ and b) the Gaussian function fitted Raman spectrum of the film in the range of 100–400 cm$^{-1}$ using the PeakFit software. The Raman mode of the Si (100) substrate is located at 523.5 cm$^{-1}$ and the solid line in Fig. 7a serves as a visual guide.

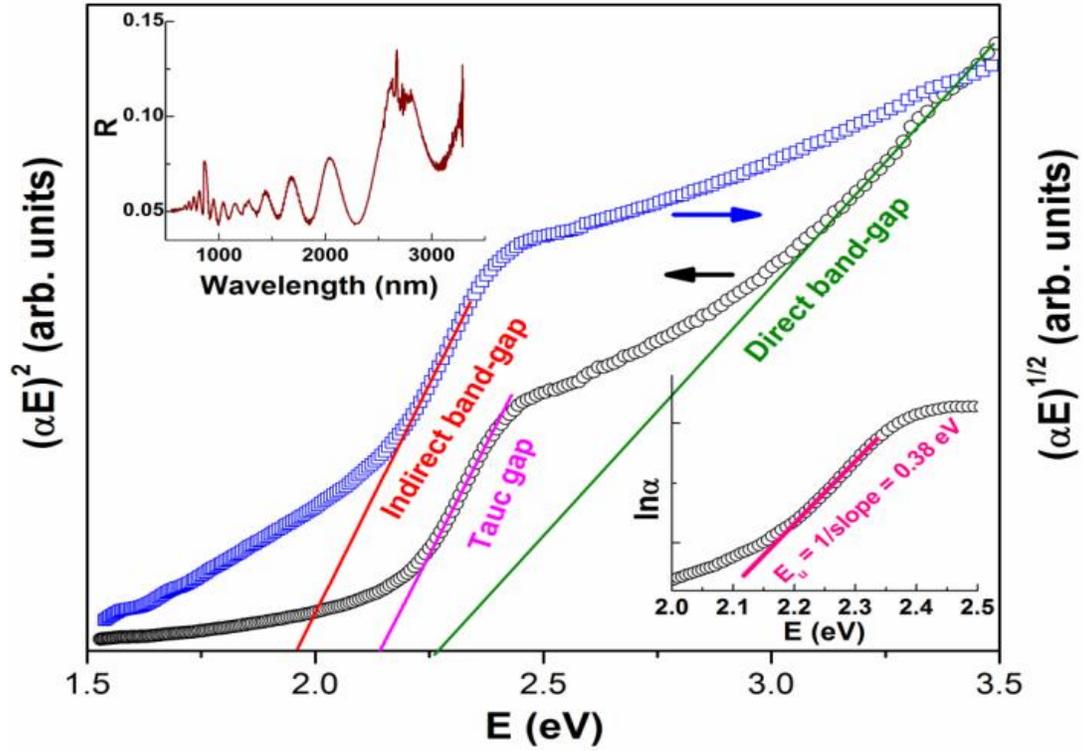

**Fig. 8** The plots of $(\alpha E)^2$ (left x-axis) and $(\alpha E)^{1/2}$ (right x-axis) as a function of photon energy ($E$) for the BiFeO$_{2.85}$ film. The top and bottom insets display the reflectance ($R$) versus wavelength and $ln\alpha$ versus $E$, respectively. The solid lines in the figures represent the corresponding linear fits.

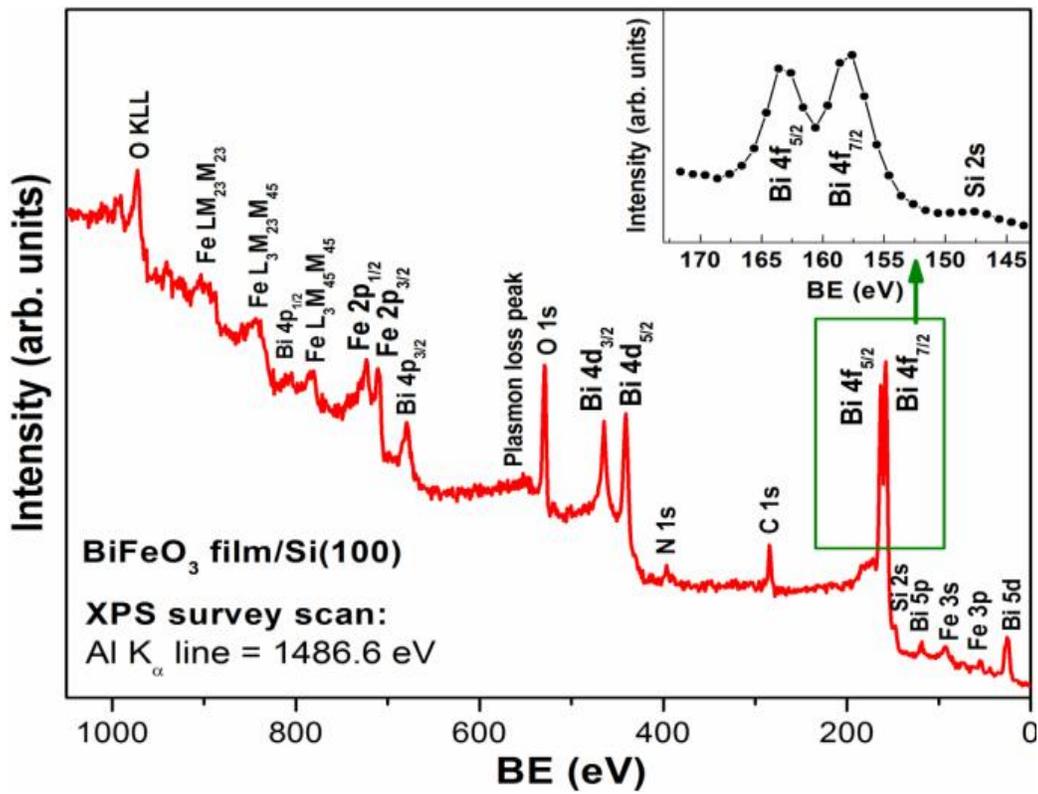

**Fig. 9** The recorded XPS surface survey scan of the BiFeO$_{2.85}$ thick film in the energy range of 0–1050 eV. The inset displays the Bi 4*f* peak of the sample in the range of 140–175 eV.

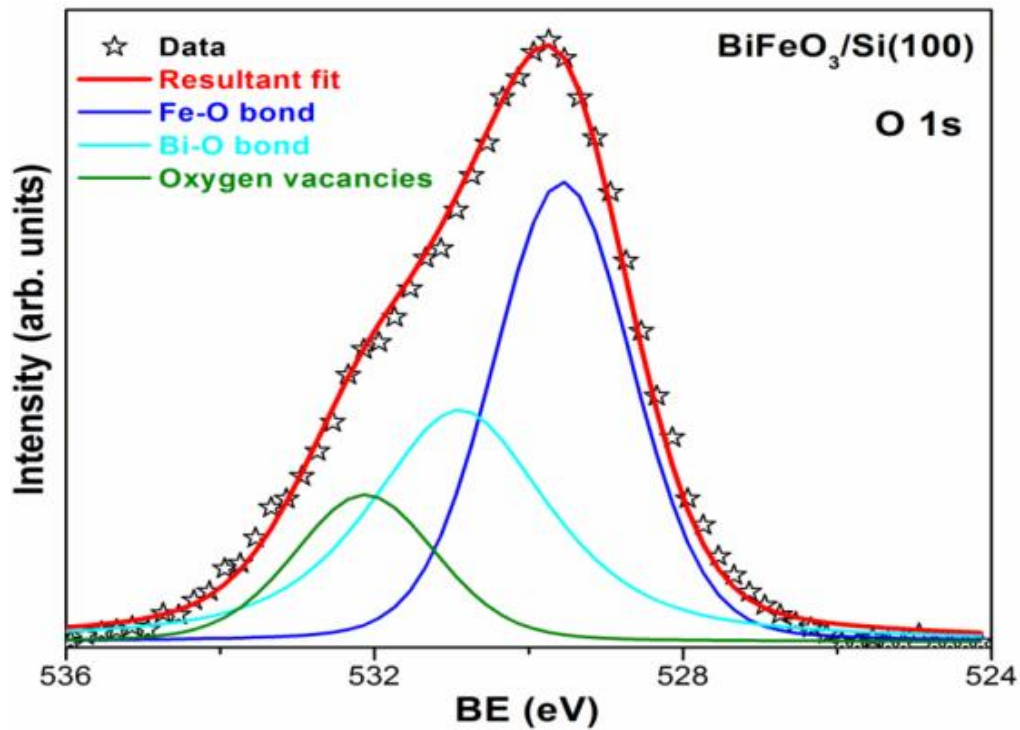

**Fig. 10** The narrow-scan XPS O 1*s* spectrum of the BiFeO$_{2.85}$ film. The open star symbol and red-colored solid line represent the experimental data and the resultant fit, respectively. The solid blue-, cyan-, and olive-colored lines correspond to the contribution of the Fe-O bond, Bi-O bond, and oxygen vacancy to the O 1*s* peak, respectively.

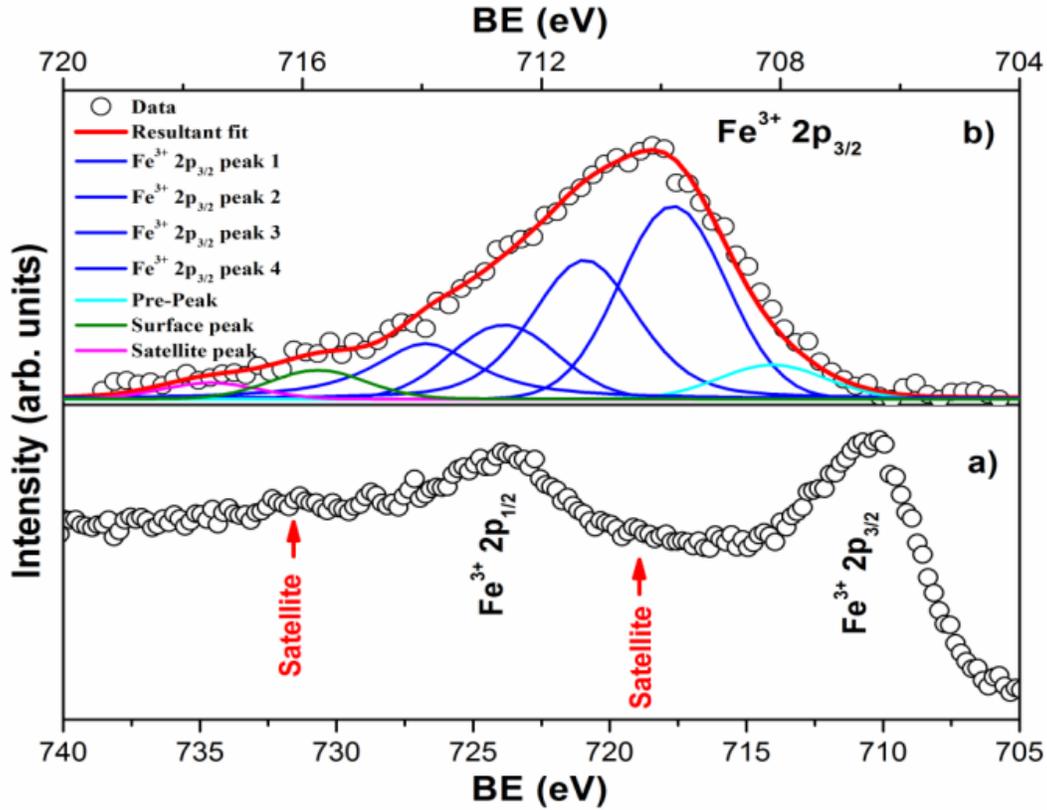

**Fig. 11** a) The narrow-scan core-level Fe 2p spectrum of the BiFeO$_{2.85}$ film; and b) the fitted Fe$^{3+}$ 2p$_{3/2}$ spectrum after the Shirley-type background subtraction. The open circular symbol and red-colored lines represent the experimental data and resultant fit, respectively. The olive-, cyan-, and magenta-colored solid lines represent the pre-peak, surface peak, and Fe$^{3+}$ 2p$_{3/2}$ satellite peak, respectively. The four blue-colored solid lines belong to the Fe$^{3+}$ 2p$_{3/2}$ multiplet peaks of the BiFeO$_{2.85}$ film.

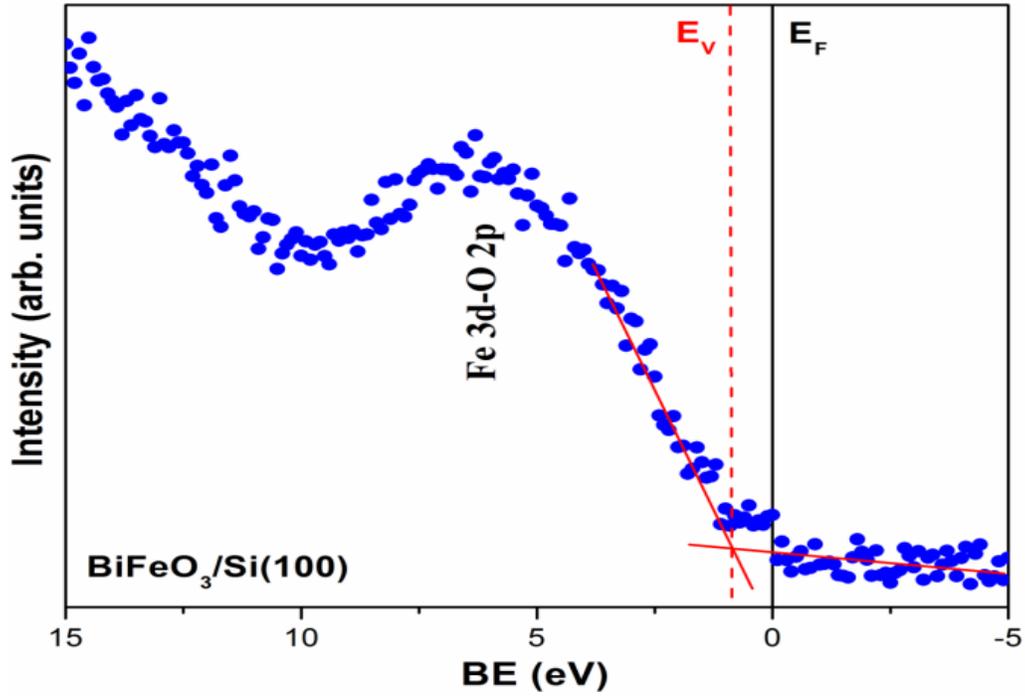

**Fig. 12** The valence band XPS spectrum of the BiFeO$_{2.85}$ film with the Fermi energy ($E_F$) marked at 0 eV and the red-colored dotted line indicating its valence band energy level ($E_V$).

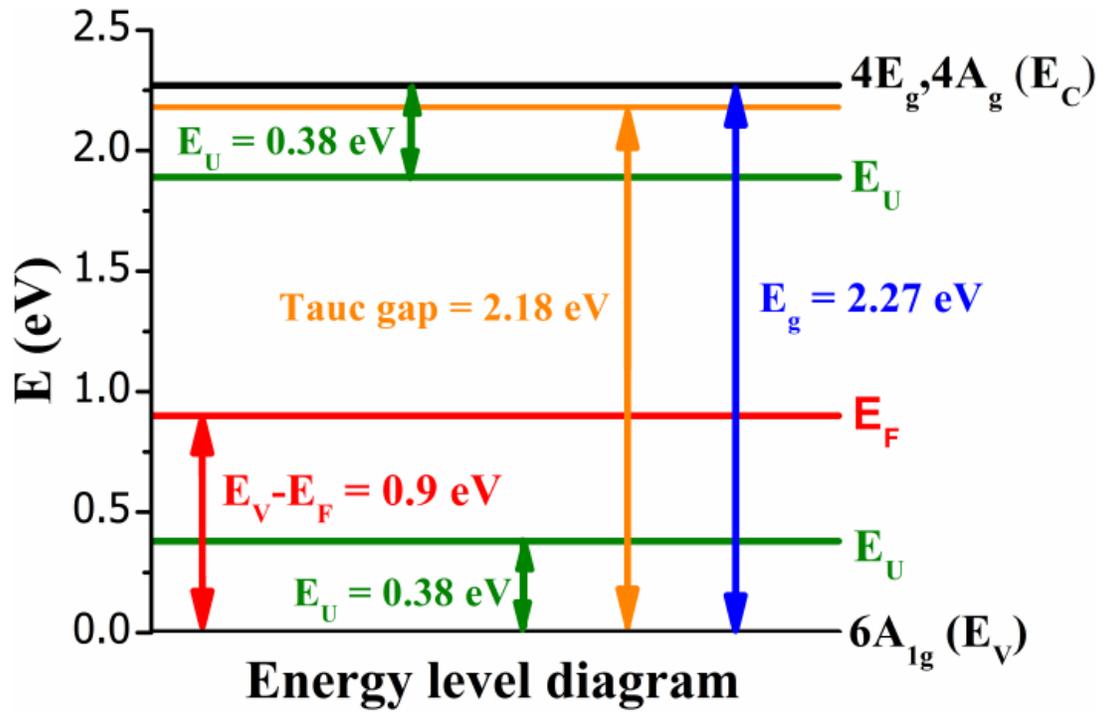

**Fig. 13** The electronic energy-level band diagram for the PLD-grown BiFeO$_{2.85}$ thick film. The symbols $E_C$, $E_V$, $E_U$, and $E_F$ correspond to its conduction band, valence band, Urbach energy, and Fermi energy, respectively.